\documentclass[twocolumn,showpacs,superscriptaddress]{revtex4-1}
\usepackage{graphicx}
\usepackage{units}
\usepackage{multirow}
\usepackage{bigstrut}
\usepackage{overpic}
\usepackage{array}
\usepackage{color}
\usepackage{subfigure}
\usepackage{amsmath}
\usepackage{float}
\usepackage[displaymath, mathlines]{lineno}
\newcommand{\doneplus}{D_1(2420)^{+}}
\newcommand{\doneminus}{D_1(2420)^{-}}
\newcommand{\done}{D_1(2420)}

\newcommand{\dplus}{D^{+}}
\newcommand{\dminus}{D^{-}}
\newcommand{\dzerobar}{\bar{D}{}^{0}}
\newcommand{\dzero}{D^{0}}
\newcommand{\ee}{e^{+}e^{-}}

\newcommand{\ddpipi}{\dplus\dminus\pip\pim}
\newcommand{\pip}{\pi^+}

\newcommand{\pim}{\pi^-}

\newcommand{\mev}{\,\rm{MeV}}

\newcommand{\mevcc}{\,\rm{MeV}/c^2}
\newcommand{\gev}{\,\rm{GeV}}

\newcommand{\gevcc}{\,\rm{GeV}/\it{c}^{\rm{2}}}

\begin{document}

\title{\boldmath Study of $e^{+}e^{-}  \to  \dplus \dminus \pip \pim$ at center-of-mass energies from 4.36 to 4.60 GeV  }

\author{
\begin{small}
\begin{center}
M.~Ablikim$^{1}$, M.~N.~Achasov$^{10,d}$, P.~Adlarson$^{59}$, S. ~Ahmed$^{15}$, M.~Albrecht$^{4}$, M.~Alekseev$^{58A,58C}$, A.~Amoroso$^{58A,58C}$, F.~F.~An$^{1}$, Q.~An$^{55,43}$, Y.~Bai$^{42}$
% ...
\\
\vspace{0.2cm}
(BESIII Collaboration)\\
\vspace{0.2cm} {\it
$^{1}$ Institute of High Energy Physics, Beijing 100049, People's Republic of China\\
% ...
$^{4}$ Bochum Ruhr-University, D-44780 Bochum, Germany\\
% ...
$^{10}$ G.I. Budker Institute of Nuclear Physics SB RAS (BINP), Novosibirsk 630090, Russia\\
% ...
$^{15}$ Helmholtz Institute Mainz, Johann-Joachim-Becher-Weg 45, D-55099 Mainz, Germany\\
% ...
$^{42}$ Southeast University, Nanjing 211100, People's Republic of China\\
$^{43}$ State Key Laboratory of Particle Detection and Electronics, Beijing 100049, Hefei 230026, People's Republic of China\\
% ...
$^{55}$ University of Science and Technology of China, Hefei 230026, People's Republic of China\\
% ...
$^{58}$ (A)University of Turin, I-10125, Turin, Italy; (B)University of Eastern Piedmont, I-15121, Alessandria, Italy; (C)INFN, I-10125, Turin, Italy\\
$^{59}$ Uppsala University, Box 516, SE-75120 Uppsala, Sweden\\
% ...
$^{d}$ Also at the Novosibirsk State University, Novosibirsk, 630090, Russia\\
% ...
}\end{center}
\end{small}
}

\date{\small \today}

\begin{abstract}
We report a study of the $e^{+}e^{-} \to  D^+ D^- \pi^+ \pi^-$ process using $e^{+}e^{-}$ collision data samples with an integrated luminosity of $2.5\,\rm{fb}^{-1}$ at center-of-mass energies from 4.36 to 4.60 GeV, collected with the BESIII detector at the BEPCII storage ring.
The $D_{1}(2420)^+$ is observed in the $D^+ \pi^+ \pi^-$ mass spectrum.
The mass and width of the $D_{1}(2420)^+$  are measured to be $(2427.2\pm 1.0_{\rm stat.}\pm 1.2_{\rm syst.})$ MeV/$c^2$ and $(23.2\pm 2.3_{\rm stat.} \pm2.3_{\rm syst.})$ MeV, respectively.
In addition, the Born cross sections of the $e^+e^- \to D_{1}(2420)^+D^- + c.c. \to D^+ D^- \pi^+ \pi^-$ and  $e^+e^- \to \psi(3770) \pi^+ \pi^- \to D^+ D^- \pi^+ \pi^-$ processes are measured as a function of the center-of-mass energy.
\end{abstract}


\maketitle


\section{Introduction}

Recent discoveries of charmonium-like states that do not fit naturally with the predicted charmonium states in the quark model have stirred up great experimental and theoretical interests~\cite{Olsen:2017bmm,Guo:2017jvc,Karliner:2017qhf,Liu:2019zoy,Brambilla:2019esw}. 
Among these so-called $XYZ$ states, the observations of the $Y(4260)$\cite{2005142001} and $Z_c(4430)$\cite{2008142001} states have drawn special attention, and stimulated extensive discussions on their structures.
Some calculations indicate that the $Y(4260)$ is possibly a $\done \bar{D}$ molecular state, while the $Z_c(4430)$ is possibly a $\done \bar{D}{}^{*} $ molecular state~\cite{132003,0708,Ma:2014zua,Chen:2016qju}. Hence, more studies on the properties of the involved $D_1(2420)$, such as mass and width, are helpful to better understand the nature of these exotic candidate states. 

The lightest charmonium state above the $D \bar{D}$ threshold is the $\psi(3770)$ resonance, which is considered to have the quantum numbers of $1^{3}D_{1}$~\cite{pdg2019,Eichten:203}. 
Its spin-triplet partner $1^{3}D_{2}$ candidate, $X(3823)$, has been observed in the process $e^{+}e^{-} \to X(3823)\pip \pim$ at BESIII~\cite{011803}. 
Analogously, it is interesting to study the production of the $\psi(3770)$ in the process $e^{+}e^{-} \to \psi(3770) \pip \pim$~\cite{114029}, 
which is observed  at $\sqrt{s}=$4415.6 MeV at BESIII~\cite{BAM204}. More precise measurements at different energy points are desired, as it provides an important way to investigate the intrinsic nature of the $Y(4360)$ and $\psi(4415)$ by studying the transitions between these charmonium(-like) states, such as $Y(4360) \to \psi(3770) \pip \pim$ and $\psi(4415) \to \psi(3770) \pip \pim$. 

In this analysis, we study the process $e^{+}e^{-} \to \dplus \dminus \pip \pim$ at the center-of-mass (c.m.) energies, $E_{\rm c.m.}$,
from 4358.3 to 4599.5 \mev,  as listed in Table~\ref{tab:results}.
Compared to the process $e^{+}e^{-} \to \dzero \dzerobar \pip \pim$, this final state has the advantage of being free from $D^*$ intermediate states, which greatly simplifies the analysis.  
We reconstruct the $\dplus$ via its high branching fraction decay $K^{-} \pip \pip$ and adopt a recoil-mass technique to identify the $D^-$ and related resonant states. Unless explicitly mentioned otherwise, inclusion of charge conjugate mode is implied throughout the context.  
Clear signals of the $\doneplus$ and $\psi(3770)$ are extracted in this data set via their decays to $\dplus \pip \pim$ and $\dplus \dminus$, respectively.
The resonance parameters of the $\doneplus$ are measured.
Additionally, the Born cross sections of $e^{+}e^{-} \to \doneplus D^- +c.c. \to\dplus \dminus \pip \pim$ and $e^{+}e^{-} \to \psi(3770) \pip \pim\to\dplus \dminus \pip \pim$ are measured at each $E_{\rm c.m.}$.

\section{The experiment and data sets}

The BESIII detector is a magnetic spectrometer~\cite{Ablikim:2009aa} located at the Beijing Electron
Positron Collider (BEPCII)~\cite{Yu:IPAC2016-TUYA01}. The cylindrical core of the BESIII detector consists of a helium-based multilayer drift chamber (MDC), a plastic scintillator time-of-flight system (TOF), and a CsI(Tl) electromagnetic calorimeter (EMC), which are all enclosed in a superconducting solenoidal magnet providing a 1.0~T magnetic field. The solenoid is supported by an octagonal flux-return yoke with resistive plate counter muon identifier modules interleaved with steel. The acceptance of charged particles and photons is 93\% over $4\pi$ solid angle. The charged-particle momentum resolution at $1~{\rm GeV}/c$ is $0.5\%$, and the $dE/dx$ resolution is $6\%$ for the electrons from Bhabha scattering. The EMC measures photon energies with a resolution of $2.5\%$ ($5\%$) at $1$~GeV in the barrel (end cap)
region. The time resolution of the TOF barrel part is 68~ps, while that of the end cap part is 110~ps.

The $E_{\rm c.m.}$ of the seven data sets are measured using di-muon events~\cite{energyMeas}, and the corresponding luminosities are measured with large-angle Bhabha scattering events~\cite{luminosityUncertainty}.
To optimize selection criteria, estimate the detection efficiency, and understand background contributions, we simulate the $e^{+}e^{-}$ annihilation processes with the {\sc kkmc} ~\cite{KKMC} generator, which takes into account continuum processes, initial state radiation (ISR), and inclusive $D^{(*)}_{(s)}$ production.
The known decay rates are taken from the Particle Data Group (PDG)~\cite{pdg2019}, and the decays are modeled with {\sc evtgen}~\cite{Lange:2001uf}.
The remaining decays are simulated with the {\sc lundcharm} package~\cite{Chen:2000tv}.
The four-body process $\ee \to \ddpipi$ is generated considering the intermediate resonances $\ee \to \doneplus \dminus$ assuming the relative orbital angular momentum of $\doneplus$-$\dminus$ in $s$-wave, and $\ee \to \psi(3770) \pip \pim$ assuming $\psi(3770) \pip \pim$ uniformly distributed in momentum phase space, along with the subsequent decays $\doneplus \to \dplus \pip \pim$ and $\psi(3770) \to \dplus \dminus$, respectively.
We simulate one million events for each process at different $E_{\rm c.m.}$.
All simulated Monte Carlo (MC) events are processed in a {\sc geant4}-based~\cite{GEANT4} software package, taking into account detector geometry and response.

\section{Event selection and Data analysis}

\subsection{Event selections}
\begin{figure*}[tp!]
\centering
\begin{overpic}[width=0.3\linewidth]{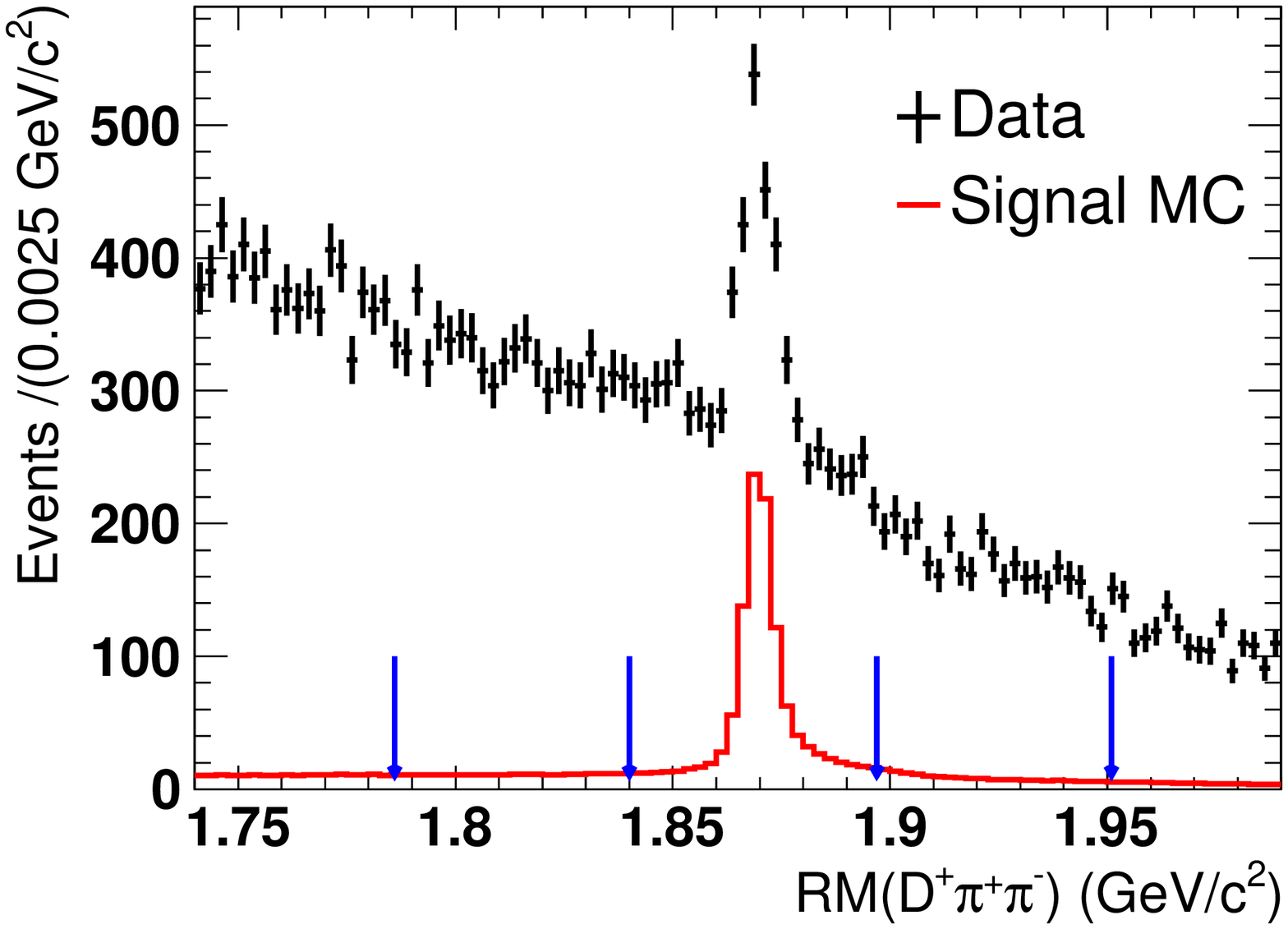}
\put(25,30){\textbf{(a)}}
\end{overpic}
\begin{overpic}[width=0.3\linewidth]{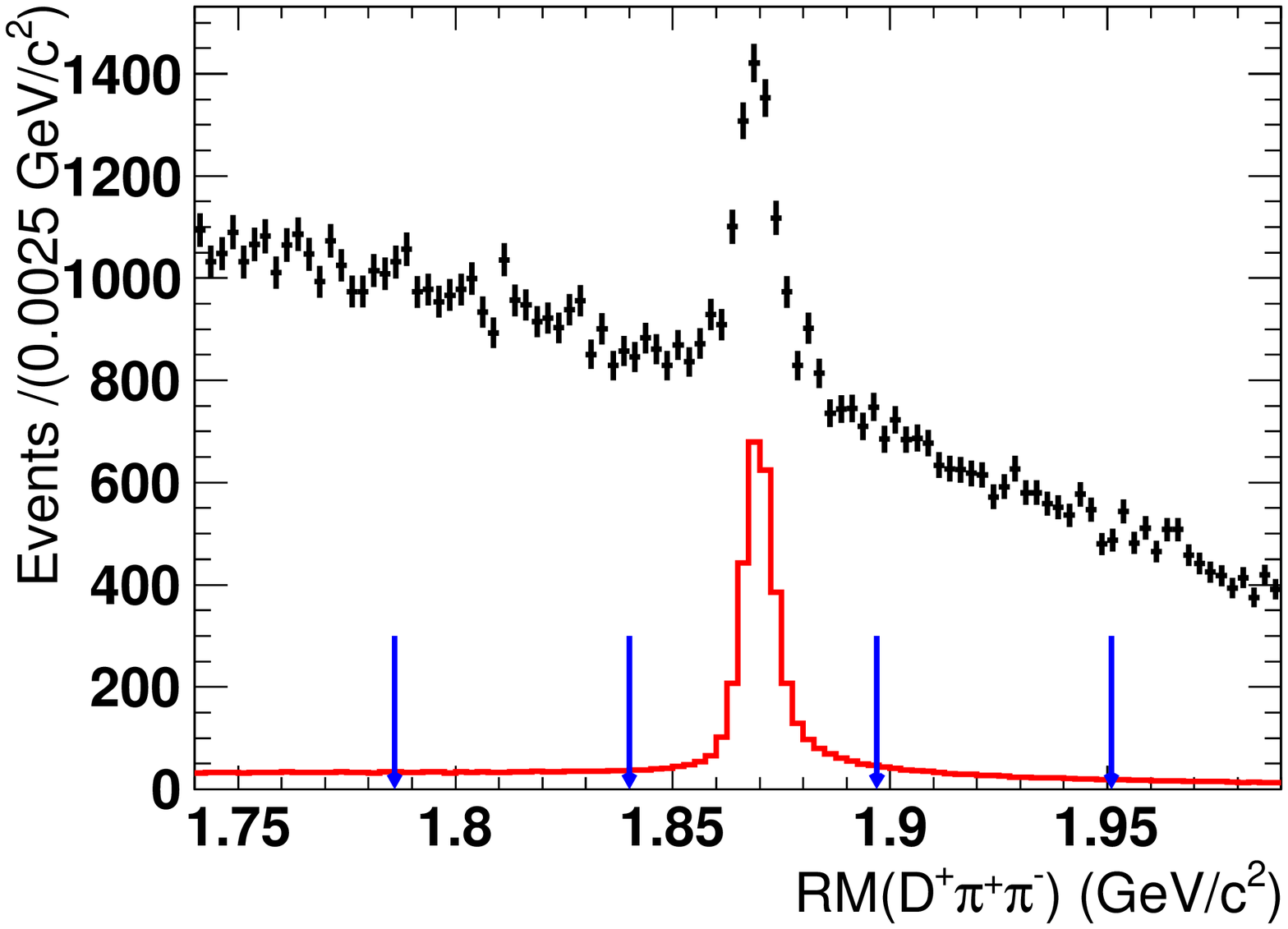}
\put(25,30){\textbf{(b)}}
\end{overpic}
\begin{overpic}[width=0.3\linewidth]{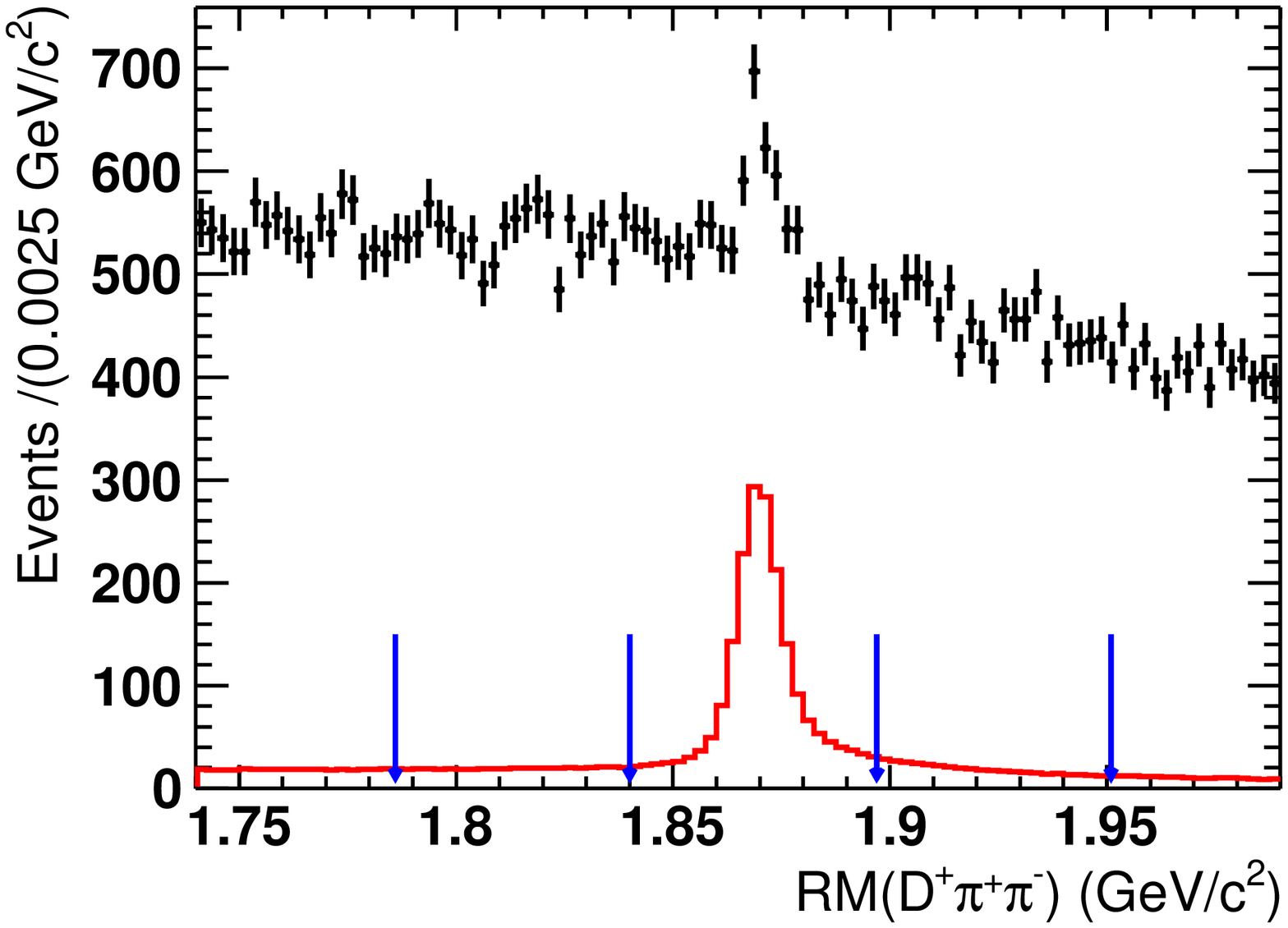}
\put(25,30){\textbf{(c)}}
\end{overpic}
\caption{(Color online) Plots (a), (b) and (c) are the recoil masses of $D^+ \pip \pim$ at $E_{\rm c.m.}=4358.3$, $4415.6$ and $4599.5\mev$, respectively. The points correspond to data and the histograms correspond to the signal MC simulations (with arbitrary normalizations). The (blue) arrows denote the sideband regions.}
\label{fig:rmdpipi}
\end{figure*}

To reconstruct the $D^+$ meson, charged track candidates for one $K^-$ and two $\pi^+$ in the MDC are selected.
For each track, the polar angle $\theta$ defined with respect to the $e^{+}$ beam is required to satisfy $|\rm{cos\theta}|<0.93$.
The closest approach to the $e^{+}e^{-}$ interaction point is required to be within $\pm 10~\rm{cm}$ along the beam direction and within $\pm 1~\rm{cm}$ in the plane perpendicular to the beam direction.
A track is identified as a $\pi(K)$ when the PID probabilities satisfy $\mathcal{P}(\pi)>\mathcal{P}(K)$ ($\mathcal{P}(K)>\mathcal{P}(\pi)$), according to the information of $\mathrm{d}E/\mathrm{d}x$ and TOF.
We reconstruct $D^+$ candidates by considering all possible combinations of the charged tracks which are required to originate from a common vertex.
The quality of the vertex fit is required to satisfy $\chi^{2}_{\rm{VF}}<100$.
We constrain the reconstructed $\dplus$ mass with a kinematic fit to the nominal $\dplus$ mass~\cite{pdg2019}, and require the fit quality $\chi^{2}_{\rm{KF}}<20$.
We then require the presence of one additional $\pip\pim$ pair, with neither track used in the reconstructed $\dplus$.  
The identification of the signal process $\ee \to \ddpipi$ is based on the recoil mass spectra of $\dplus \pip \pim$, $RM(\dplus \pip \pim)$, which are shown in Fig.~\ref{fig:rmdpipi}.
The rate of multiple candidates per event is about 10\%, and is corrected for via the MC efficiency.  

\begin{figure*}[tp!]
\centering
\begin{overpic}[width=0.3\linewidth]{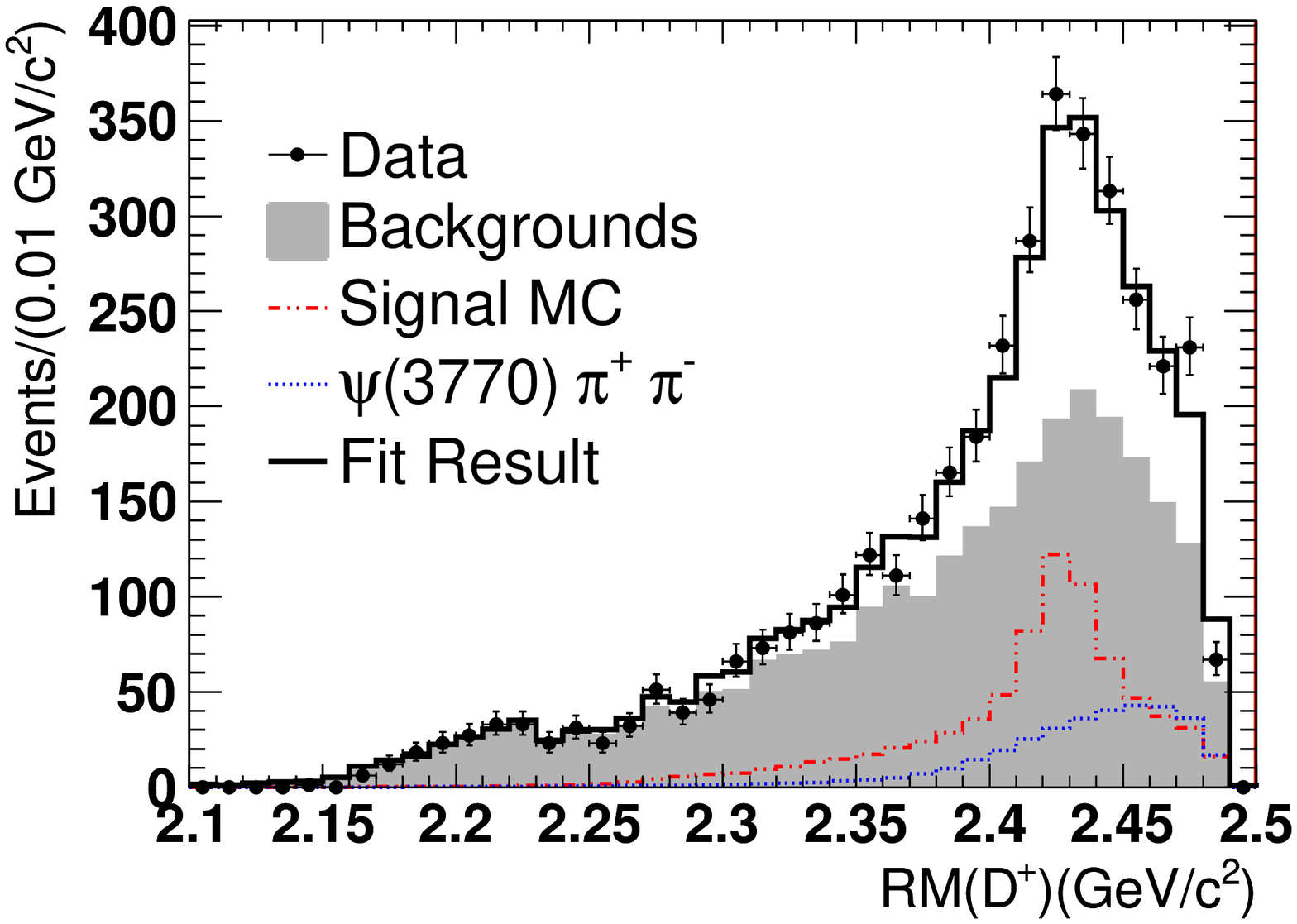}
\put(25,26){\textbf{(a)}}
\end{overpic}
\begin{overpic}[width=0.3\linewidth]{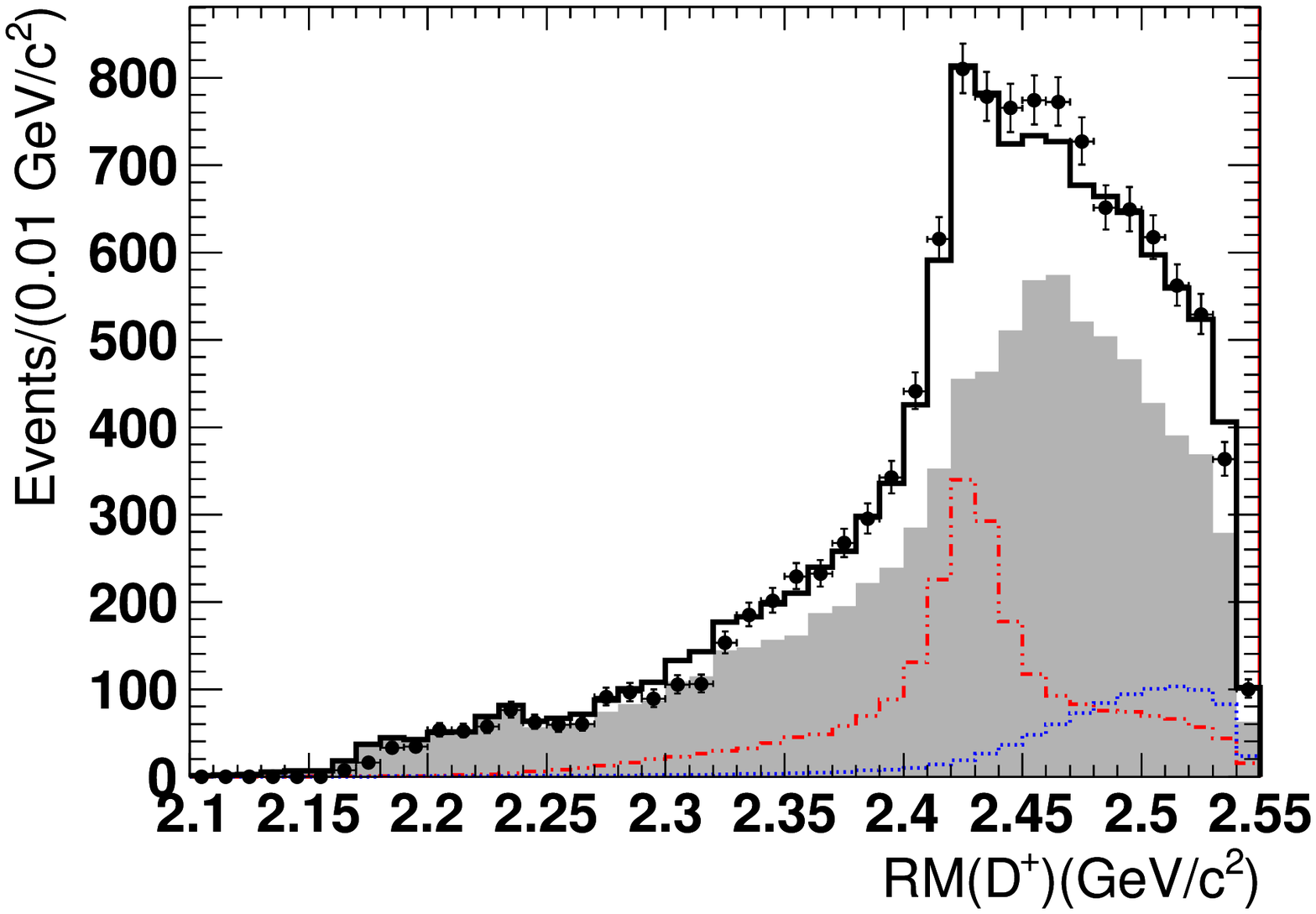}
\put(25,26){\textbf{(b)}}
\end{overpic}
\begin{overpic}[width=0.3\linewidth]{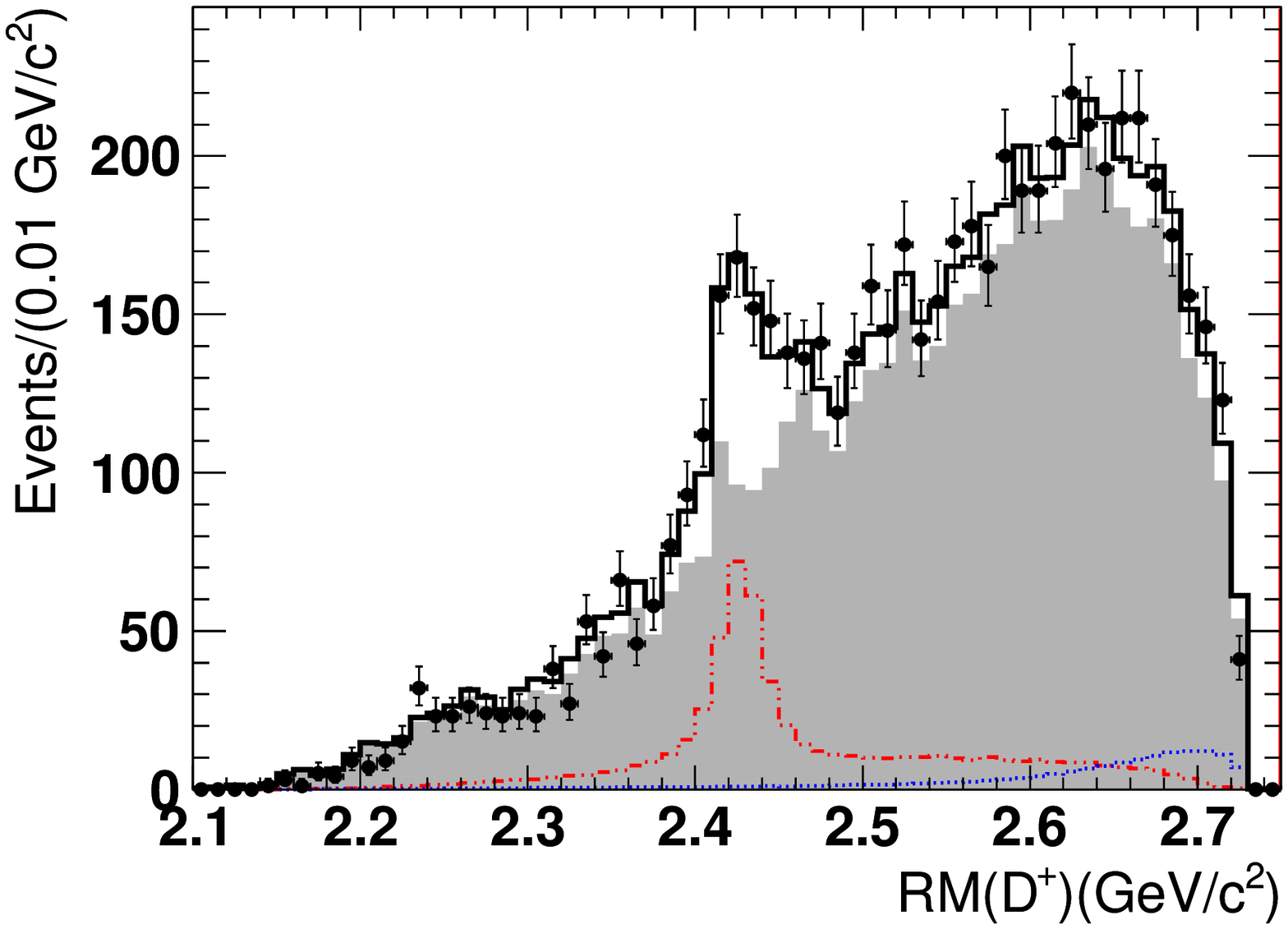}
\put(25,26){\textbf{(c)}}
\end{overpic}
\caption{(Color online) (a), (b) and (c) correspond to the simultaneous fits to the $RM(\dplus)$ distributions at $E_{\rm c.m.}=4358.3$, $4415.6$ and $4599.5\mev$, respectively. The points with error bars are data, the (gray) shaded histograms are backgrounds, the (red) dash-dotted lines are $\doneplus \dminus +c.c. \to \dplus \dminus \pip \pim$ signal process and the (blue) dotted lines are $\psi(3770) \pip \pim \to \dplus \dminus \pip \pim$. The (black) solid lines are the result of fit. }
\label{fig:simufit}
\end{figure*}

The peaks observed at $1.87\gevcc$ correspond to the $D^-$ meson signals.  
They are consistent with the MC simulations of the $D^{+} D^{-} \pip \pim$ final state.
The background contributions are due to random combinations of charged tracks.
We further restrict the candidate events to the region $1.855<RM(\dplus \pip \pim)<1.882\gevcc$, and plot the recoiling mass of the $\dplus$, $RM(\dplus)$, as shown in Fig.~\ref{fig:simufit}.
Enhancements around the $\doneplus$ nominal mass are clearly visible.
We take the events with $RM(\dplus \pip \pim)$ in the sideband regions of $(1.786, 1.840)\gevcc$ and $(1.897, 1.951)\gevcc$ which are illustrated in Fig.~\ref{fig:rmdpipi}, as samples representing the combinatorial background contributions in the distributions of $RM(D^+)$.
This approach has been verified using the corresponding distributions of the background contributions from the inclusive MC samples. 
It is found that the sideband samples correctly reproduce the background in the signal region of $RM(D^{+}\pip\pim)$.
Besides the contributions from $\doneplus\dminus$, there is a clear excess of the data over background contributions from the sideband at high $RM(\dplus)$ mass.  
It is consistent with being from the process $\ee\to\psi(3770) \pip \pim\to\dplus \dminus \pip \pim$.

\begin{figure*}[tp!]
\centering
\begin{overpic}[width=0.4\linewidth]{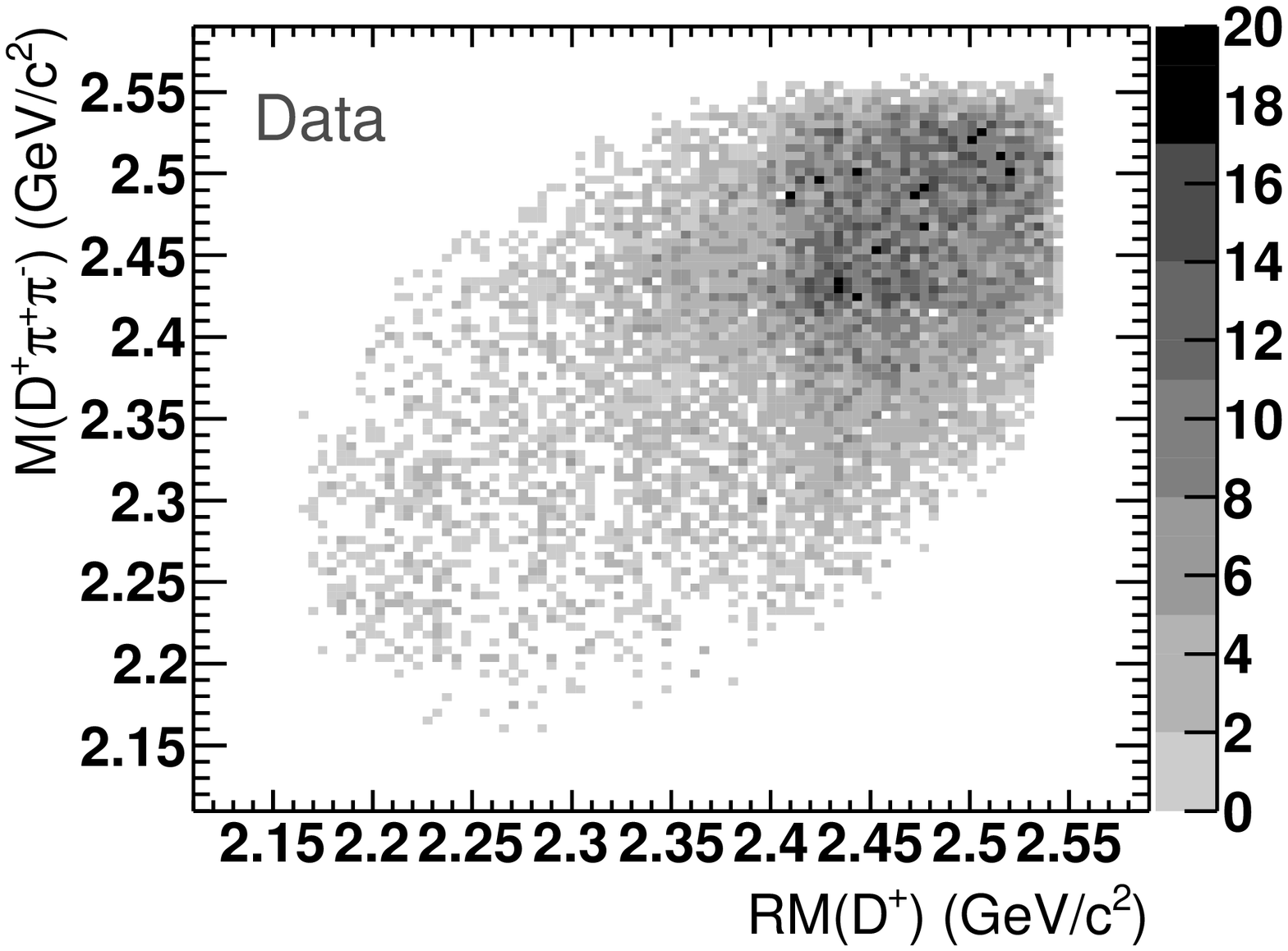}
\put(22,21){\textbf{(a)}}
\end{overpic}
\begin{overpic}[width=0.4\linewidth]{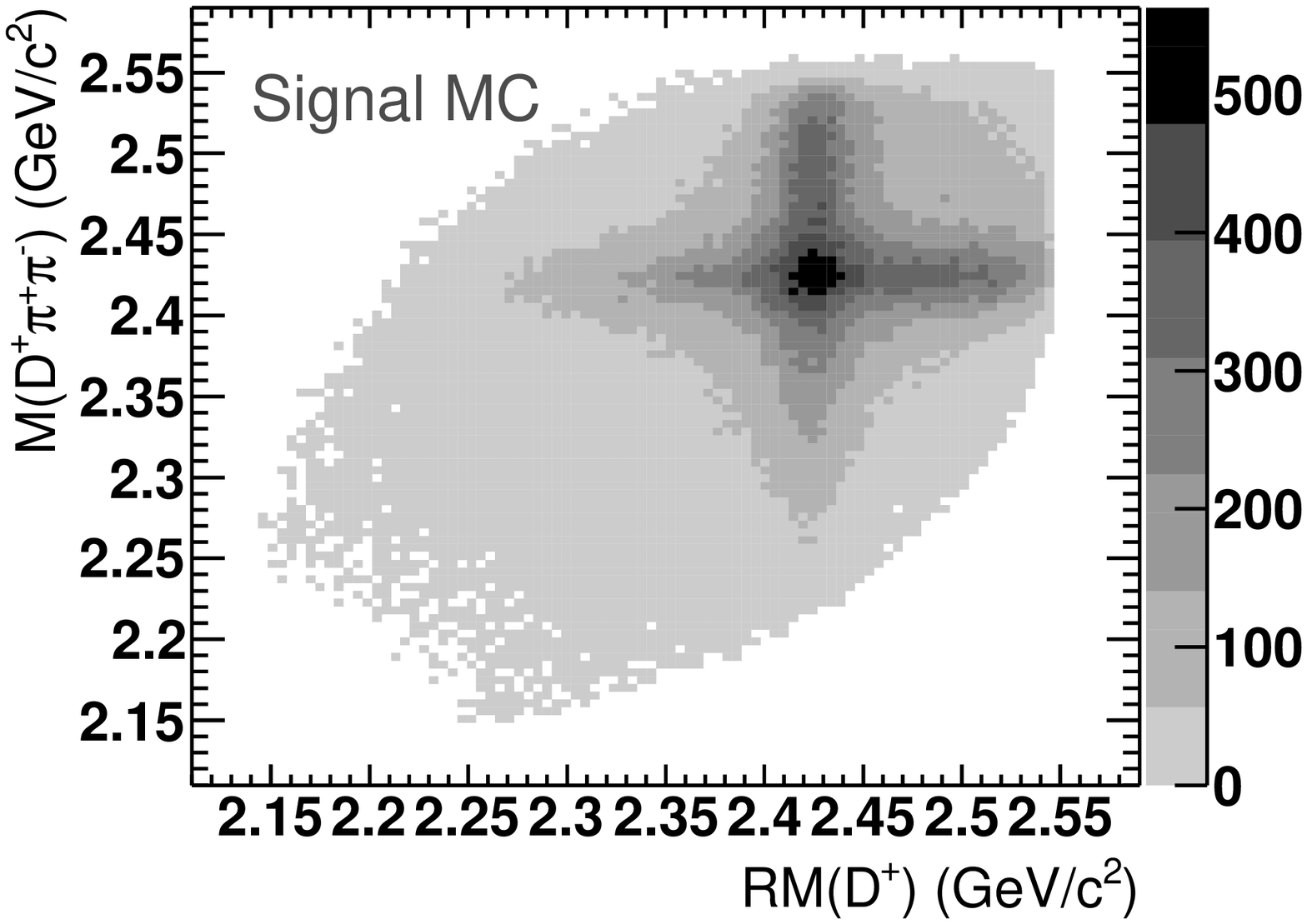}
\put(22,21){\textbf{(b)}}
\end{overpic}
\caption{ Plots (a) and (b) correspond to the scatter plot of  $M(\dplus\pip\pim)$ versus $RM(\dplus)$ in  data and $D_{1}^+D^-+c.c.$ signal MC samples at $E_{\rm c.m.}= 4415.6 \mev$, respectively. }
\label{fig:signalshape}
\end{figure*}

\subsection{Signal extraction}

The 2-dimensional distributions of $M(\dplus\pip\pim)$ versus $RM(\dplus)$ for the $\doneplus \dminus$ are shown in Fig.~\ref{fig:signalshape}. The vertical band corresponds to the $\doneminus$ signal
and the horizontal band corresponds to the $\doneplus$.  
The projection to the $RM(\dplus)$ axis (Fig.~\ref{fig:simufit}) consists of a prominent $\doneminus$ peak
and a corresponding broad bump. The contributions of $\doneplus \dminus$ and $\psi(3770) \pip \pim$ in the 
selected data are determined using fits to the $RM(\dplus)$ one-dimensional distribution.
The shape of this distribution is described using templates obtained from the signal MC simulation.
In order to perform a likelihood scan of the resonance parameters, we generate a series of $\doneplus$ signal MC with different values of mass and width, and smear these template shapes with a Gaussian function to take into account the resolution difference between data and MC simulations.
The width of the Gaussian function is fixed to the difference of resolution in $RM(D^+)$ for the control sample of $\ee \to \dplus \dminus$.
The signal shape for the mode $\psi(3770) \pip \pim$ is obtained from the MC simulation, 
where the resonance parameters of the $\psi(3770)$ are taken from the PDG~\cite{pdg2019}. The relativistic Breit-Wigner function~\cite{pdg2019} is used to model  the resonance lineshape of the $\psi(3770)$ and $\doneminus$.

A simultaneous unbinned maximum likelihood fit to the data samples is performed at three high luminosity energy points of $E_{\rm c.m.}=4358.3, 4415.6$ and $4599.5\mev$, with the resonance parameters of the $\doneplus$ in common for all fits.  
The shapes and magnitudes of the combinatorial backgrounds are fixed according to the sample of the sideband events in $RM(\dplus\pip\pim)$, while the magnitudes of the $\doneplus\dminus$ and $\psi(3770)\pip\pim$ are the free parameters of the fit.
The sum of the fitting components is shown in Fig.~\ref{fig:simufit}.
We obtain the mass and width of the $\doneplus$ to be $(2427.2\pm1.0)\mevcc$ and $(23.2\pm2.3)\mev$, respectively.
The signal yields are also measured, as listed in Table~\ref{tab:results}.
Here, the contribution of the non-resonant four-body process $\ee\to D^+ D^- \pip\pim$ is neglected in the fit, as an alternative fit including this process gives its size consistent with zero.

In addition, we analyze the data samples at $E_{\rm c.m.}=4487.4, 4467.1, 4527.1$ and $4574.5\mev$ with relatively low luminosities.
We apply the same strategy to extract the signal yields of the $\doneplus\dminus$ and $\psi(3770)\pip\pim$, except that we fix the resonance parameters for the $\doneplus$ according to the aforementioned fit results.

\subsection{Cross section measurement}
\begin{figure*}[tp!]
\centering
\begin{overpic}[width=0.45\linewidth]{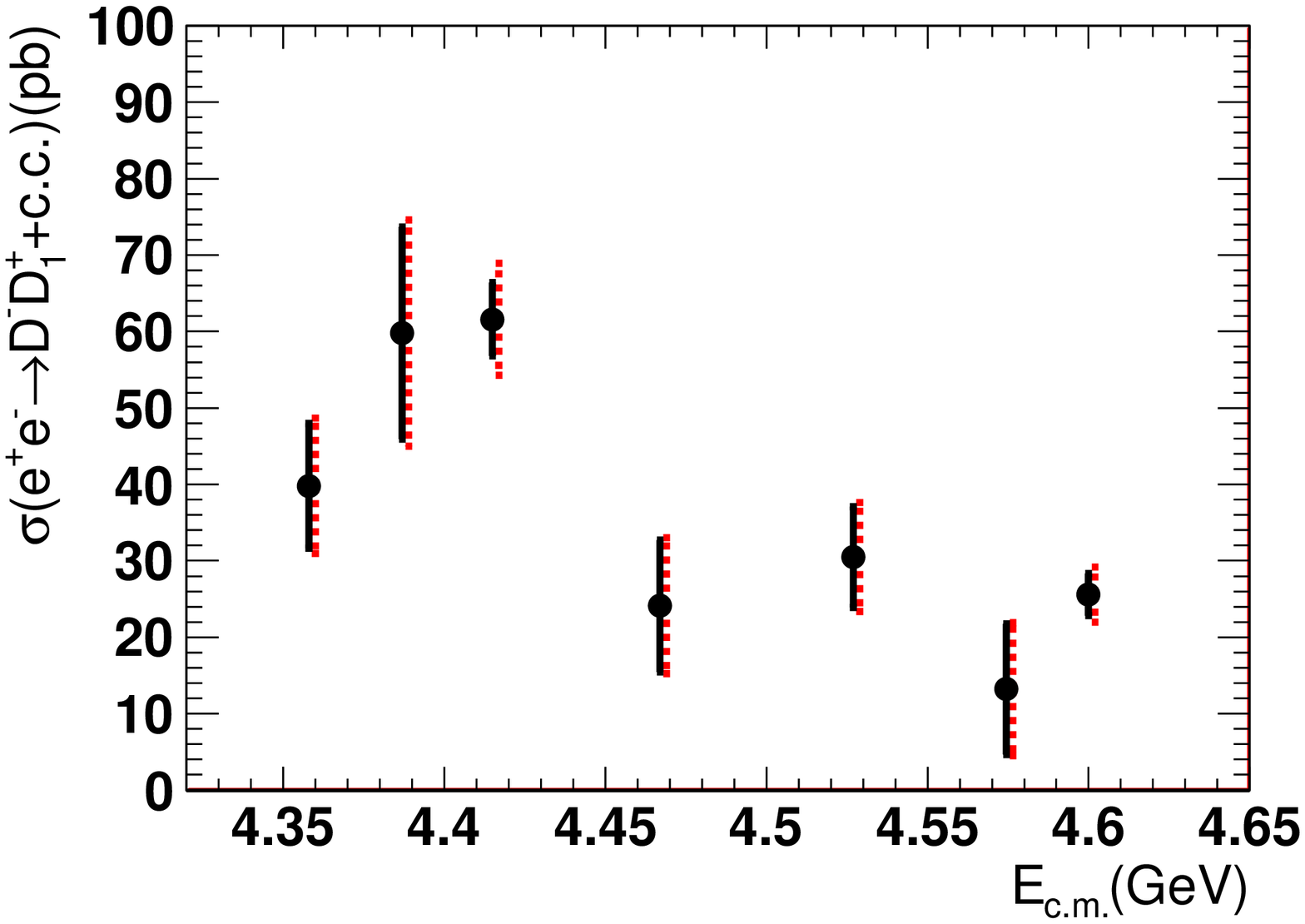}
\put(70,60){\textbf{(a)}}
\end{overpic}
\begin{overpic}[width=0.45\linewidth]{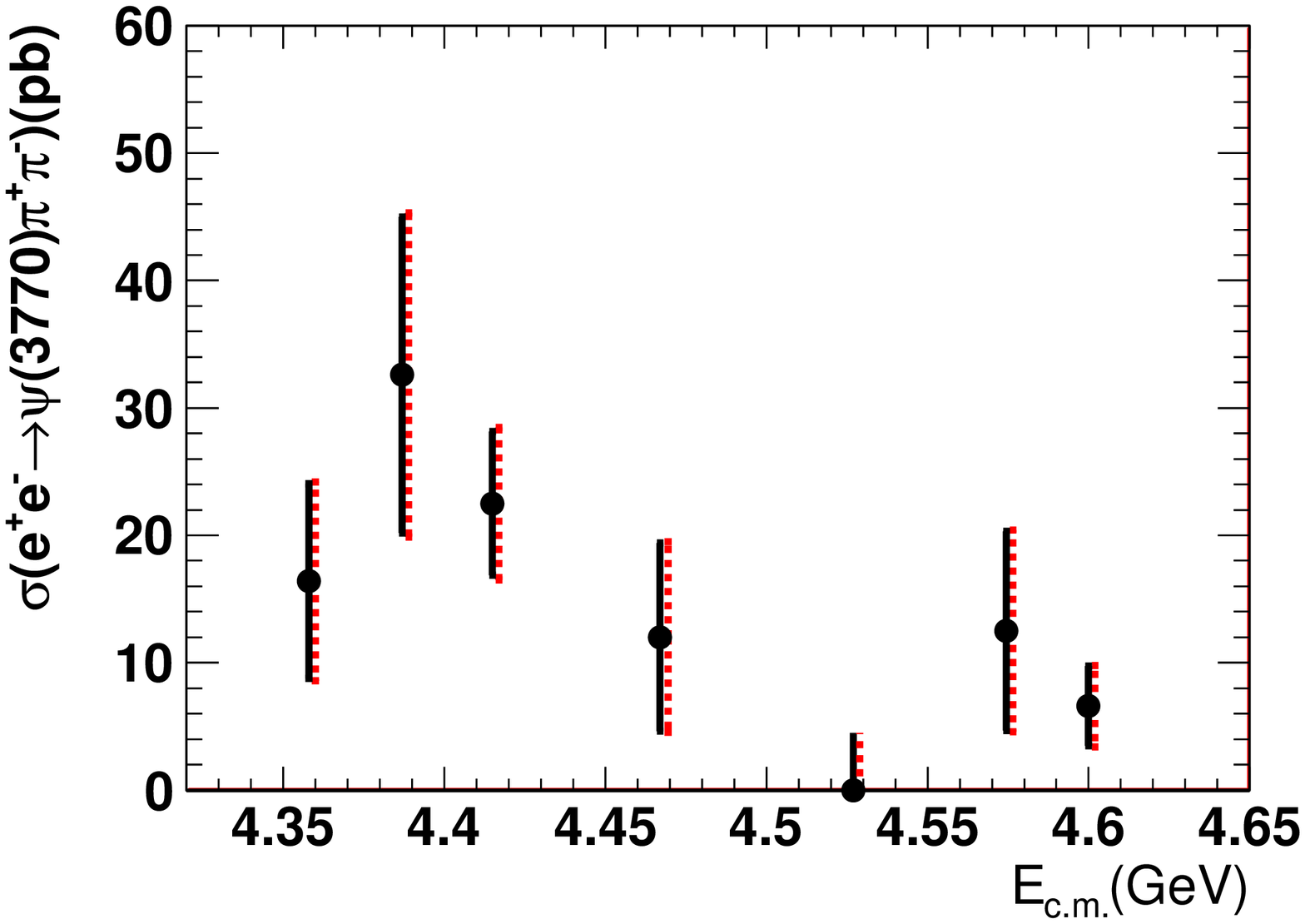}
\put(70,60){\textbf{(b)}}
\end{overpic}
\caption{(Color online) The measured Born cross sections of the signal processes (a) $e^{+}e^{-} \to \doneplus \dminus + c.c. \to \dplus \dminus \pip \pim$and  (b) $e^{+}e^{-} \to \psi(3770) \pip \pim \to \dplus \dminus \pip \pim$. The (black) solid lines are the sum of statistical uncertainties and independent systematic uncertainties in quadrature, the (red) dot lines are total uncertainties.}
\label{fig:crosssection}
\end{figure*}

The Born cross section is calculated with
\begin{equation}
\sigma_i = \frac{ n_i^{\rm sig}} { 2 \mathcal{L}\mathcal{B}\varepsilon_i  (1+\delta_i^\text{rad}) \frac{1}{|1-\Pi|^2}},
\label{eq:bornxs}
\end{equation}
where index $i$ denotes the respective signal process, $n_i^{\rm sig}$ is the observed signal yield, $\mathcal{L}$ is the integrated luminosity, $\mathcal{B}$ is the branching fraction ${\mathcal B}(\dplus \to K^- \pip\pip) = (9.38\pm0.16)\%$ ~\cite{pdg2019}, $\varepsilon_i$ is the detection efficiency,
$(1+\delta_i^\text{rad})$ is the radiative correction factor which is obtained from a QED calculation using the line shape of the data cross section of signal process as input in an iterative procedure, and $ \frac{1}{|1-\Pi|^2}$ is the vacuum polarization factor~\cite{585}.
The trigger efficiencies for the two processes are 100\%, as there are at least 5 charged tracks detected~\cite{Berger:triggeff}.
The processes $\ee\to\doneplus\dminus +c.c. \to\dplus\dminus\pip\pim$ and $\ee\to\psi(3770)\pip\pim\to\dplus\dminus\pip\pim$ are denoted with index $i=1$ and $i=2$, respectively.
The calculated Born cross sections are given in Table~\ref{tab:results} and plotted in Fig.~\ref{fig:crosssection}.
We evaluate the statistical significance by the ratio of the maximum likelihood value and the likelihood value for a fit with a null-signal
hypothesis.
For the energy points with low statistical significances, we determine the upper limits for the cross sections which are calculated by using the signal yield upper limits $n^{\rm UL}$ in Eq.~\eqref{eq:bornxs}. 
The upper limit $n^{\rm UL}$ at  90\% confidence level is obtained with a Bayesian approach scanning the expected signal yield. 
The probability is calculated from the Gaussian-smeared likelihood to take into account the systematic uncertainty.

\begin{table*}[th!]
  \begin{center}
     \caption{The numbers  relevant to the Born cross section measurements, where the first uncertainties are statistical, the second are independent systematic uncertainties, and the third are common systematics. The index of $1$ represents the process $e^{+}e^{-} \to \doneplus \dminus +c.c. \to \dplus \dminus \pip \pim$ while the index of $2$ represents the process $e^{+}e^{-} \to \psi(3770) \pip \pim \to \dplus \dminus \pip \pim$. The upper limits correspond to the 90\% confidence level.  The symbol $\mathcal{S}$ refers to the statistical significance. }
  \begin{tabular}{c|ccccccc}
      \hline \hline
        \footnotesize{$E_{\rm c.m.}$(MeV)} &$\mathcal{L}(\rm pb^{-1})$    &\footnotesize{$n_1^{\rm sig}$}    &\footnotesize{$\varepsilon_{1}(\%)$}  &\footnotesize{$1+\delta_{1}^{\rm rad}$} &\footnotesize{$\frac{1}{|1-\Pi|^2}$}    &\footnotesize{$\sigma_1(\rm pb)$}    &$\mathcal{S}_{1}$  \\ \hline
        \footnotesize{4358.3}            &\footnotesize{543.9}                &\footnotesize{$810\pm109$}          &\footnotesize{23.90}            &\footnotesize{0.795}                      &\footnotesize{1.051}                   &\footnotesize{$39.8\pm5.3\pm6.2\pm3.5$}     	&\footnotesize{7.9$\sigma$} 
        \\
        \footnotesize{4387.4}            &\footnotesize{55.6}                 &\footnotesize{$125\pm28$}             &\footnotesize{23.20}            &\footnotesize{0.822}                      &\footnotesize{1.051}                   &\footnotesize{$59.8\pm13.3\pm3.9\pm5.3$}          &\footnotesize{4.9$\sigma$}  
        \\
        \footnotesize{4415.6}            &\footnotesize{1090.7}             &\footnotesize{$2454\pm111$}         &\footnotesize{22.56}             &\footnotesize{0.820}                     &\footnotesize{1.053}                   &\footnotesize{$61.6\pm2.8\pm3.9\pm5.5$}            &\footnotesize{24.9$\sigma$} 
        \\
        \footnotesize{4467.1}            &\footnotesize{111.1}               &\footnotesize{$100\pm28$}            &\footnotesize{20.92}             &\footnotesize{0.904}                    &\footnotesize{1.055}                   &\footnotesize{$24.1\pm6.6\pm5.6\pm2.1$}             &\footnotesize{3.9$\sigma$}   
        \\
        \footnotesize{4527.1}            &\footnotesize{112.1}               &\footnotesize{$122\pm24$}           &\footnotesize{19.27}              &\footnotesize{0.935}                    &\footnotesize{1.055}                   &\footnotesize{$30.5\pm5.9\pm3.0\pm2.7$}             &\footnotesize{5.8$\sigma$}  
         \\
        \footnotesize{4574.5}            &\footnotesize{48.9}                 &\footnotesize{$24\pm15$($<43$)}            &\footnotesize{18.22}               &\footnotesize{1.029}                   &\footnotesize{1.055}                    &\footnotesize{$13.2\pm8.3\pm2.2\pm1.2$($<23.7$)}            &\footnotesize{1.7$\sigma$}  
        \\
        \footnotesize{4599.5}            &\footnotesize{586.9}                &\footnotesize{$572\pm56$}         &\footnotesize{17.92}               &\footnotesize{1.075}                   &\footnotesize{1.055}                   &\footnotesize{$25.6\pm2.5\pm1.2\pm2.3$}             &\footnotesize{11.7$\sigma$} 
        \\
        \hline\hline
    \end{tabular}
     \label{tab:results}
  \end{center}
  \end{table*}
  \begin{table*}[th!]
  \begin{center}
    \begin{tabular}{c|ccccccc}
      \hline \hline
        \footnotesize{$E_{\rm c.m.}$(MeV)}  &$\mathcal{L}(\rm pb^{-1})$    &\footnotesize{$n_2^{\rm sig}$}    &\footnotesize{$\varepsilon_{2}(\%)$}  &\footnotesize{$1+\delta_{2}^{\rm rad}$}  &\footnotesize{$\frac{1}{|1-\Pi|^2}$}  &\footnotesize{$\sigma_2(\rm pb)$}  &$\mathcal{S}_{2}$ \\ \hline
        \footnotesize{4358.3}            &\footnotesize{543.9}                &\footnotesize{$323\pm101$}      &\footnotesize{23.51}                &\footnotesize{0.780}                   &\footnotesize{1.051}                   &\footnotesize{$16.4\pm5.1\pm5.7\pm1.5$}		&\footnotesize{3.8$\sigma$}
         \\
        \footnotesize{4387.4}            &\footnotesize{55.6}                 &\footnotesize{$66\pm24$($<97$)}         &\footnotesize{23.43}                &\footnotesize{0.789}                   &\footnotesize{1.051}                   &\footnotesize{$32.6\pm12.0\pm3.1\pm2.9$($<47.8$)}		&\footnotesize{2.9$\sigma$}
        \\
        \footnotesize{4415.6}            &\footnotesize{1090.7}              &\footnotesize{$900\pm97$}      &\footnotesize{22.51 }               &\footnotesize{0.826}                   &\footnotesize{1.053}                   &\footnotesize{$22.5\pm2.4\pm5.1\pm2.0$} 		&\footnotesize{10.3$\sigma$}
         \\
        \footnotesize{4467.1}            &\footnotesize{111.1}                &\footnotesize{$50\pm27$($<88$)}        &\footnotesize{19.78}                &\footnotesize{0.960}                   &\footnotesize{1.055}                  &\footnotesize{$12.0\pm6.4\pm3.7\pm1.1$($<21.1$)} 		&\footnotesize{1.9$\sigma$}
        \\
        \footnotesize{4527.1}            &\footnotesize{112.1}                &\footnotesize{$0^{+20}_{-0}$($<30$)}              &\footnotesize{17.21}                 &\footnotesize{1.151}                   &\footnotesize{1.055}                   &\footnotesize{$0^{+4.5}_{-0}$($<6.8$)}					&\footnotesize{$-$}
        \\
        \footnotesize{4574.5}            &\footnotesize{48.9}                 &\footnotesize{$23\pm14$($<44$)}       &\footnotesize{15.39}                &\footnotesize{1.236}                   &\footnotesize{1.055}                   &\footnotesize{$12.5\pm7.8\pm0.7\pm1.1$($<23.9$)}		&\footnotesize{1.7$\sigma$}
        \\
        \footnotesize{4599.5}            &\footnotesize{586.9}                &\footnotesize{$152\pm 58$($<227$)}     &\footnotesize{14.93}              &\footnotesize{1.319}                   &\footnotesize{1.055}                   &\footnotesize{$6.6\pm2.5\pm1.9\pm0.6$($<9.9$)}		&\footnotesize{2.7$\sigma$}
        \\
        \hline\hline
    \end{tabular}
  \end{center}
  \end{table*}

\section{Systematic uncertainties}

The systematic uncertainties of the measurement of the $\doneplus$ resonance parameters and the Born cross sections listed in Tables~\ref{tab:sys_cs} and~\ref{tab:sys_cs_2} include correlated (common) contributions, from tracking, PID, luminosity measurements, vacuum polarization factors, interference effect and the input branching fraction, as well as uncorrelated (independent) contributions from background shapes, mass scaling, detector resolution, signal shape due to the angular distributions, and radiative corrections.  
\begin{table*}[th!]
  \begin{center}
      \caption{Summary of systematic uncertainties on the $\doneplus$ resonance parameters and the Born cross sections for the high luminosity energy points.}
  \begin{tabular}{cl|cc|ccc|ccc}
      \hline \hline
   &\multirow{2}{*}{Source} & \multirow{2}{*}{$ m$\footnotesize{($\mevcc$)}}  & \multirow{2}{*}{$\Gamma$\footnotesize{($\mev$)}}  & \multicolumn{3}{c|}{$\sigma_{1}$(\%)} & \multicolumn{3}{c}{$\sigma_{2}$(\%)} \\ \cline{5-10}
   & & & &  \footnotesize{4358.3\,MeV} &   \footnotesize{4415.6\,MeV} &  \footnotesize{4599.5\,MeV} 
   &  \footnotesize{4358.3\,MeV} &   \footnotesize{4415.6\,MeV} &  \footnotesize{4599.5\,MeV}   \\ \hline 
  
      \multirow{6}{*}{Common} & Tracking            &     &      & 5.0 & 5.0 & 5.0& 5.0 & 5.0 & 5.0  \\
      &Particle ID          &     &      & 5.0 & 5.0 & 5.0& 5.0 & 5.0 & 5.0  \\
      &Luminosity            &     &      & 1.0 & 1.0 & 1.0& 1.0 & 1.0 & 1.0  \\
      &Vacuum polarization  &     &      & 0.1 & 0.1 & 0.1& 0.1 & 0.1 & 0.1   \\
      &Interference  &     &      & 5.0 & 5.0 & 5.0 & 5.0 & 5.0 & 5.0   \\
      &Input $\mathcal{B}$  &     &      & 1.7 & 1.7 & 1.7& 1.7 & 1.7 & 1.7\\ 
      \cline{2-10}
           &   Sum &   &      & 8.9 & 8.9 & 8.9 & 8.9 &8.9 &8.9 \\
      \hline      
      \multirow{5}{*}{Independent}&Background           & 0.1 & 0.6  & 3.8 & 2.3 & 2.6& 2.1 & 3.3 & 14.1  \\
      &Mass scale            & 0.8 &      &     &     &    &     &     &   \\
      &Detector resolution  & 0.1 & 1.5  & 1.5 & 0.7 & 0.8& 2.8 & 1.6 & 0.3  \\
      &Angular distribution  &0.9 &1.6 & 15.0 & 4.9 &3.1 & 34.1 & 22.2 & 25.1  \\
      &Radiative correction &     &      & 2.4 & 3.1 & 2.1& 2.5 & 2.5 & 1.8   \\
                     \cline{2-10}
           &   Sum &  1.2 &  2.3    & 15.7  & 6.3  & 4.6 & 34.6 & 22.6 & 28.8 \\        \hline
            Total&            & 1.2 & 2.3  & 18.1 & 10.9 & 10.0 &35.7  & 24.3 &30.2   \\
        \hline\hline
    \end{tabular}
    \label{tab:sys_cs}
  \end{center}
\end{table*}
\begin{table*}[th!]
  \begin{center}
      \caption{Summary of systematic uncertainties on the Born cross sections for the low luminosity energy points. The total systematic uncertainty is taken as the quadratic sum of the individual uncertainties.}
  \begin{tabular}{ll|cccc|cccc}
      \hline \hline
      \multirow{2}{*}{Source} & &  \multicolumn{4}{c|}{$\sigma_{1}$(\%)}  & \multicolumn{4}{c}{$\sigma_{2}$(\%)}  \\ \cline{3-10}
     & & \footnotesize{4387.4\,MeV} &   \footnotesize{4467.1\,MeV} &  \footnotesize{4527.1\,MeV}  &  \footnotesize{4574.5\,MeV}  
      & \footnotesize{4387.4\,MeV} &   \footnotesize{4467.1\,MeV} &  \footnotesize{4527.1\,MeV}  &  \footnotesize{4574.5\,MeV}  \\ \hline
      Common &      & 8.9 & 8.9& 8.9& 8.9&8.9 &8.9 &8.9 &8.9\\
      \hline
      \multirow{4}{*}{Independent}& Backgrounds          	     & 5.3 & 6.9 & 8.1& 6.0 & 4.2 & 13.1&3.3 &5.2 \\
      & Detector resolution   	     & 0.7 & 0.7 & 0.6& 0.6 & 1.6 & 1.6  & 0.3&0.3 \\
      & Angular distribution 	     & 2.7 & 22.1 &5.2&15.8&7.6 & 28.0 & 0.9 &0.9 \\
      & Radiative correction 	    & 2.5 & 2.4 & 2.4& 2.1 & 3.2 & 2.8&3.9 &1.9   \\   
        \cline{2-10}
           &   Sum &   6.5 &  23.3  & 9.9 & 17.0 & 9.4 & 31.1 & 5.2 & 5.6   \\
      \hline  
          Total &                          & 11.0 & 24.9 &13.3 &19.2 &12.9 & 32.3 & 10.3 & 10.5   \\
        \hline\hline
    \end{tabular}
    \label{tab:sys_cs_2}
  \end{center}
\end{table*}
\begin{itemize}
\item  Uncertainties of tracking and PID are each $1\%$ per track~\cite{BESIII112005}.
\item  The systematic uncertainties due to background contributions are estimated by leaving their magnitudes free in the fit and changing the ranges of the sideband regions. The statistical errors of the sideband samples are also included in the background uncertainty.
\item  The mass scale uncertainty for $\doneplus$ mass is estimated from the mass shift of $RM(\dplus)$ in the control sample of $\ee \to \dplus \dminus$.
To be conservative, the largest mass shifts among the three high luminosity energy points, $0.8\mevcc$, is assigned as the systematic uncertainty due to the mass scale.
\item  The uncertainties due to the detector resolution are accounted for by changing the Gaussian widths for smearing the signal shape in the fit to the $RM(\dplus)$ distribution.
These widths, representing the resolution difference between data and MC, are varied within the uncertainty obtained from the control sample of $\ee \to \dplus \dminus$ events.
The resultant maximum changes on the numerical results are considered as the systematic uncertainties due to the detector resolution.
\item The uncertainty of modeling the angular distributions of the signal processes  are studied by repeating the analysis procedure on the basis of new signal model. For  $\ee \to \doneplus \dminus$, we considered two extreme cases of $1+\cos^2 \theta_{D_1}$ and $1-\cos^2 \theta_{D_1}$, where $\theta_{D_1}$ is the helicity angle of the $\doneplus$ in the rest frame of the initial $\ee$ system. For  $\ee \to \psi(3770) \pip \pim$, a model, named as JPIPI~\cite{Lange:2001uf} in {\sc evtgen}, is considered. The maximum changes on the results are taken as systematic uncertainties.

\item Interference effects among the processes $\ee \to \doneplus \dminus$, $\doneminus \dplus$, and $\psi(3770) \pip \pim$ are tested by varying input parameters of the matrix elements. In the test, $\doneplus\to D_0^*(2300) \pi^+$ is assumed, as favored in Ref.~\cite{Abe:2004sm}, while $\pip\pim$ $S$-wave is assumed in $\ee\to\psi(3770) \pip \pim$. The average relative sizes of the interference effects are taken into account as systematic uncertainties.
 
\item  The uncertainty of luminosity measurement is $1\%$, as given in Ref.~\cite{luminosityUncertainty}.
\item  The uncertainty of radiative correction is calculated by using the generator {\sc kkmc}. Initially, the observed signal events are assumed to originate from the $Y(4260)$ resonance to obtain the efficiency and ISR correction factor.
Then, the measured line shape is used as input to calculate the efficiency and ISR correction factor again.
This procedure is repeated until the difference between the subsequent iterations is comparable with the statistical uncertainty.
We take the difference of the radiative correction factors between the last two iterations as the systematic uncertainty.
\item  We take $0.1\%$ as the uncertainty of the vacuum polarization  factor, which is calculated in Ref.~\cite{585}.
\item The input branching fraction of $\dplus\to K^-\pip\pip$ in PDG has the relative uncertainty of 1.7\%, which is taken into account.
\end{itemize}

The systematic uncertainties are summarized in Tables~\ref{tab:sys_cs} and~\ref{tab:sys_cs_2}; 
the sum of different uncertainties are obtained by adding up all the relevant contributions in quadrature.

\section{Discussion and summary}

In summary, based on $\ee$ annihilation data at $E_{\rm c.m.}=4358.3$, $4387.4$, $4415.6$, $4467.1$,  $4527.1$, $4574.5$, and $4599.5\mev$, we studied the $\doneplus$ in the mass spectrum of $\dplus \pip \pim$ system 
in the final state of $e^{+}e^{-} \to \dplus \dminus \pip \pim$.
The mass and width of  the $\doneplus$ are measured to be $(2427.2\pm 1.0 \pm1.2)\mevcc$ and $(23.2\pm2.3 \pm2.3) \mev$, respectively, which are consistent with the corresponding world-average values of  $(2423.2\pm2.4)\mevcc$ and $(25\pm6)\mev$ in PDG~\cite{pdg2019} and  have better precisions.
More accurate resonance parameters of the $\doneplus$ will better control the uncertainties of theoretical calculations for the $D_1(2420)\bar{D}$  and $D_1(2420)\bar{D}^{*}$ molecular explanations for the $Y(4260)$ and $Z_c(4430)$ states, respectively.

The Born cross sections of $e^{+}e^{-} \to \doneplus \dminus + c.c. \to \dplus \dminus \pip \pim$  and $e^{+}e^{-} \to \psi(3770) \pip \pim \to \dplus \dminus \pip \pim$ are measured as functions of the center-of-mass energy.
The cross section line shape is consistent with previous BESIII measurement based on full reconstruction method~\cite{BAM204}.
There are some indications of enhanced cross sections for both processes between 4.36 and $4.42 \gev$, where the reported states $Y$(4360) and $\psi(4415)$ locate. Hence, the measured cross sections can be useful inputs to the properties of these states.

\section{Acknowledgments}

The BESIII collaboration thanks the staff of BEPCII and the IHEP computing center for their strong support. 
This work is supported in part by National Key Basic Research Program of China under Contract No. 2015CB856700; National Natural Science Foundation of China (NSFC) under Contracts Nos. 11625523, 11635010, 11735014, 11805064, 11822506; National Natural Science Foundation of China (NSFC) under Contract No. 11835012; the Chinese Academy of Sciences (CAS) Large-Scale Scientific Facility Program; Joint Large-Scale Scientific Facility Funds of the NSFC and CAS under Contracts Nos. U1532257, U1532258, U1732263, U1832207; CAS Key Research Program of Frontier Sciences under Contracts Nos. QYZDJ-SSW-SLH003, QYZDJ-SSW-SLH040; 100 Talents Program of CAS; INPAC and Shanghai Key Laboratory for Particle Physics and Cosmology; German Research Foundation DFG under Contract Nos. Collaborative Research Center CRC 1044, FOR 2359; Istituto Nazionale di Fisica Nucleare, Italy; Koninklijke Nederlandse Akademie van Wetenschappen (KNAW) under Contract No. 530-4CDP03; Ministry of Development of Turkey under Contract No. DPT2006K-120470; National Science and Technology fund; The Swedish Research Council; U. S. Department of Energy under Contracts Nos. DE-FG02-05ER41374, DE-SC-0010118, DE-SC-0012069; University of Groningen (RuG) and the Helmholtzzentrum fuer Schwerionenforschung GmbH (GSI), Darmstadt.


\end{document}